# Simultaneous measurement of anisotropic thermal conductivity and thermal boundary conductance of 2-dimensional materials

Mizanur Rahman[1], Mohammadreza Shahzadeh[2], Simone Pisana[1,2]

1. Department of Physics and Astronomy, York University, Toronto, Canada
2. Department of Electrical Engineering and Computer Science, York University, Toronto, Canada

**Abstract**

The rapidly increasing number of 2-dimensional (2D) materials that have been isolated or synthesized provides an enormous opportunity to realize new device functionalities. Whereas their optical and electrical characterization have been more readily reported, quantitative thermal characterization is more challenging due to the difficulties with localizing heat flow. Optical pump-probe techniques that are well-established for the study of bulk materials or thin-films have limited sensitivity to in-plane heat transport, and the characterization of the thermal anisotropy that is common in 2D materials is therefore challenging. Here we present a new approach to quantify the thermal properties based on the magneto-optical Kerr effect that yields quantitative insight into cross-plane and in-plane heat transport. The use of a very thin magnetic material as heater/thermometer increases in-plane thermal gradients without complicating the data analysis in spite of the layer being optically semi-transparent. The approach has the added benefit that it does not require the sample to be suspended, providing insight of thermal transport in supported, device-like environments. We apply this approach to measure the thermal properties of a range of 2D materials which are of interest for device applications, including single layer graphene, few-layer h-BN, single and few layer $MoS_2$, and bulk $MoSe_2$ crystal. The measured thermal properties will have important implications for thermal management in device applications.

**Introduction**

Research on graphene and other 2-dimensional (2D) materials is intense due to their exceptional physical, electrical, and thermal properties, creating opportunities to realize new device functionalities as well as enhancing current technologies[1–5]. This class of materials also provides great flexibility in creating complex heterostructures by combining distinct layers into one stack[1,2]. The thermal and electrical properties of the resulting heterostructures can be controlled for a specific application as these materials cover a wide range of properties.

Graphene is superior due to its excellent thermal properties, fast electron mobility, and highest mechanical strength, and is also extremely common in Van der Waals heterostructures[3]. These exceptional properties make graphene attractive for a broad range of applications in electronics and optoelectronics devices, circuits, photonics to biosensors and solar cells[1–5]. Hexagonal boron nitride (h-BN), also referred to as white graphene, has a similar structure and high thermal conductivity but it is an electrical insulator. h-BN is very convenient as a dielectric and tunnel barrier, as it can be readily integrated with other 2D materials to create heterostructures[1]. As a substrate for graphene devices, h-BN significantly enhances the charge carrier mobility and thermal conductivity of graphene which are desirable for high-performance graphene devices[1,6,7]. At room temperature, the thermal conductivity of h-BN supported graphene is calculated to be 1,347 W/mK, about twice that of measured $SiO_2$-supported single layer graphene[7].

Semiconducting layered materials based on transitional metal dichalcogenides (TMDs) such as $MoS_2$, $WS_2$, and $MoSe_2$ have also shown attractive qualities for applications in electronic, optoelectronic, and spintronic devices[8–13]. $MoS_2$ is particularly attractive in photonics and optoelectronics given the large on-off current ratios in devices and layer-dependent bandgap, which is direct for single-layer $MoS_2$, indirect otherwise[8–10]. Monolayer $MoS_2$ exhibits a stronger photoluminescence compared to other TMDs. However, $MoSe_2$ has a higher electrical conductivity as well as a direct bandgap, which is beneficial to applications such as transistors and photodetectors[11–13].

In spite of the numerous studies that focus on the optical and electrical properties of these 2D materials, the thermal characterization has lagged behind, due to the difficulties with localizing heat flow within the material of interest and obtaining quantitative results. The thermal conductivity of these materials is highly anisotropic because of the strong atomic interactions along the basal plane compared to weak bonding present cross-plane (typically the crystal's *c*-axis). Raman thermometry has been used to measure in-plane heat transport of 2D materials[14,15], but the technique is prone to inaccuracies due to errors in accurately determining the optical absorption of the sample and heat transfer at the sample boundaries, the latter being more challenging to quantitatively assess and taken into account[15]. Additionally, Raman thermometry requires that the sample is suspended, providing limited insight of substrate-induced or interfacial effects on the thermal transport. As a result, despite the relative simplicity of the Raman technique, a more accurate approach is required, ideally capable of also measuring interfacial heat transport or transport along different directions. The optical pump-probe techniques known as time domain thermoreflectance[16–18] and frequency domain thermoreflectance[19–25] are well established to directly measure the thermal conductivity and thermal boundary conductance (TBC) of bulk materials and thin films[26,27]. These detect changes in the surface temperature of a sample subject to an optically generated heat flux, and the results are used to extract the thermophysical properties of interest. However, these techniques are not normally sensitive to lateral heat transport, and hence the characterization of anisotropic 2D materials is very challenging. The in-plane thermal conductivity of some anisotropic materials has been measured through thermoreflectance techniques by beam offsetting[17,19,22] and varying spot sizes[18]. The experimental sensitivity needs to be further increased to measure the anisotropic thermal conductivity as well as TBC of 2D materials, especially for the case of single and few-layer structures. Liu et al. recently demonstrated how using a thin transducer enhances the sensitivity to lateral heat transport in time domain thermoreflectance (TDTR)[28]. However, in TDTR the instrumentation is comparatively expensive, and the measurements are typically modulated at frequencies below 20 MHz, setting a lower limit to the thermal penetration depth and in turn limiting the ability to measure thin samples. Frequency domain thermoreflectance (FDTR) is cost effective as it does not require an ultrafast laser or electro-optic modulators, is easier to set-up as it does not use a long mechanical delay stage, and it allows modulating the measurement over a wide range of frequencies[19,23,24].

In this work we implement frequency domain magneto-optical Kerr effect (FD-MOKE), a new approach to quantify the thermal properties of anisotropic materials based on the Kerr effect, yielding enhanced sensitivity to lateral heat transport. While in FDTR the sample is typically coated with a relatively thick metal film (50-100 nm) with large thermoreflectance coefficient (known as the transducer), in FD-MOKE a thinner magnetic film is used as a transducer and the detection of the modulated surface temperature is achieved through the Kerr effect through changes in magnetization as function of temperature $dM/dT$[28]. Reducing the thickness of the metallic layer limits the lateral heat flow within it and increases the sensitivity to lateral heat flow in the sample. Additionally, a thinner transducer has a lower thermal mass and can be used to probe heat transport over shallower depths. Reducing the transducer thickness in typical thermoreflectance approaches is not convenient, as the correct interpretation of the results either requires that all the optical energy be absorbed within an optically thick transducer, or a more complicated model is needed to account for the optical properties and absorption as function of depth. Kerr detection on the other hand is only sensitive to changes in the magnetization state of the magnetic layer, so any optical contribution from other layers are not detected and do not contribute to the measurement.

Applying FD-MOKE and beam offset FD-MOKE, we have measured the thermal conductivity and thermal boundary conductance of a range of 2D materials including single-layer graphene, few-layer h-BN, single and two-layer $MoS_2$, and bulk $MoSe_2$ crystal. The thermal properties of bulk silicon and sapphire are also measured as reference to demonstrate the validity of this technique.

**Experimental**

The basic operating principle in FDTR and FD-MOKE is to impose a sinusoidal heat flux on the sample surface and detect the resulting surface temperature oscillations. The phase shift between the heat flux and the temperature oscillations are proportional to the thermophysical properties of the sample. The main difference is that FDTR samples the surface temperature through thermoreflectance ($dR/dT$), whereas FD-MOKE captures the thermally-induced changes in magnetization ($dM/dT$). The experimental setup of FD-MOKE as illustrated in figure 1 is similar to that of FDTR[19,24] but with a different detection scheme. Briefly, a sinusoidally modulated pump beam and continuous wave probe beam are combined and focused on the sample surface. Non-polarizing optics are used to route the beams in order to avoid affecting the strength of the Kerr signal. A Wollaston prism splits the beam into two orthogonal polarization states, and a half-wave plate balances their intensities before reaching a balanced photodetector. We offset the pump beam with respect to the probe using a 12 mm thick fused silica plate on an actuated optical mount. Optical spot sizes with $1/e^2$ rms values of 1.4 µm are achieved through a 40X objective, which increase the signal strength and allow us to detect signals over a wide range of frequencies, up to 50 MHz. The accurate determination of the spot sizes is critical to minimize error, and this is done by profiling the beams at their focal point across the sharp sample's edge using a piezoelectric stage.

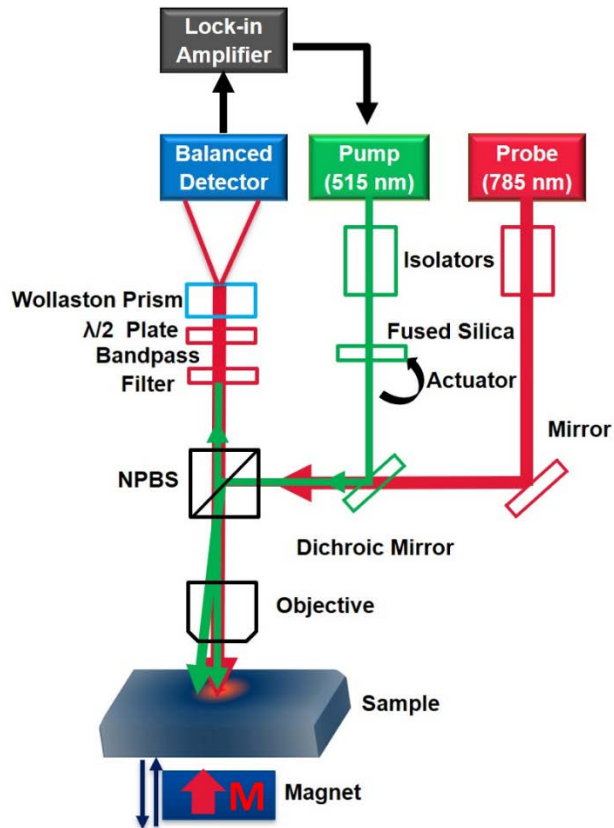

Figure 1 - Schematic diagram of the Frequency Domain Magneto-Optical Kerr Effect setup.

The optical system is set-up in the polar Kerr configuration for ease of implementation, and this implies that the detected signal is sensitive to changes of the magnetization state of the transducer in the direction perpendicular to the sample surface. In this case, transducer materials with perpendicular magnetic anisotropy would be ideal, but this would also complicate sample preparation. We choose Nickel thin films as transducer for several reasons. First, Ni is readily available and a thin film with repeatable magnetic characteristics can be easily deposited irrespective of the choice of substrate or film thickness. Since the demagnetizing field of the thin film dominates any other source of magnetic anisotropy, the remanent magnetization will be in-plane. In order to achieve Kerr contrast in the polar geometry, the magnetization needs to be brought out of the plane of the sample, so using Ni, which is a ferromagnet with relatively low magnetization ~500 emu/cc, requires a relatively weak field of ~6.3 kOe to sufficiently tilt the magnetization out of the plane. Lastly, Ni has a relatively low Curie temperature ~350 C, therefore its magnetization curve as function of temperature will have a comparatively large slope dM/dT near room temperature, aiding in the measurement of the temperature induced changes to the magnetization. A 20 nm Nickel layer serves as a transducer throughout this work, but thinner layers can also be used. This approach is suitable for characterization over a fairly broad temperature range. At high temperature the limitation is due to the reduction in signal strength, as the magnetization decreases approaching the Curie temperature. At low temperature the limitations are due to reduced signal strength, as the slope dM/dT decreases, and larger applied magnetic field requirements, since the magnetization increases and a higher field will be needed to tilt the magnetization out of plane.

To isolate the thermal phase lag of interest, we perform two measurements of the phase of the signal as function of modulation frequency. We first null the probe signal in the balanced detector by rotating the half-wave plate, then we perform the first measurement while applying a saturating perpendicular field to the sample using an external permanent magnet. This measurement is referred to as $\theta_1(f)$, and it contains the modulation frequency-dependent phase information from the temperature fluctuations in the magnetic layer, reference phase, optical phase, and electrical phase. Since the measurement was preceded by nulling the probe signal at zero applied magnetic field, only the response from the magnetic transducer contributes to the signal, whereas any other contribution such as thermoreflectance from underlying layers is rejected. Another measurement is required to determine the reference phase, and optical and electrical contributions. This measurement is referred to as $\theta_2(f)$, and is performed by detecting the pump beam. Subtracting the two frequency-dependent phases ($\theta_1 - \theta_2$) yields the desired thermal phase, which is fit to a model based on the diffusive heat equation to extract the thermal properties of interest. Two optical bandpass filters are used to separate the pump and probe beams before the balanced photodetector.

**Data modelling and sensitivity analysis**

The frequency-dependent solution to the diffusive heat equation in layered media in cylindrical coordinates is well known[16]. The temperature fluctuations detected are given by[16,17]

$$\Delta T(f) = \frac{2\pi}{A_s} \int_0^\infty G(f,k)P(k)S(k)k dk \tag{1}$$

where $P(k)$ and $S(k)$ are the Hankel transforms of the intensity profiles of the Gaussian pump and probe beams, $A_s$ is the total amplitude of the signal, and $G(f,k)$ is the Hankel transform of the frequency domain solution of the heat equation in multilayered media, with $k$ being the Hankel transform variable. $G(f,k)$ is obtained iteratively using the thermophysical properties of all the layers comprising the sample, namely the thickness $t$, volumetric heat capacity $C$, and cross-plane and in-plane thermal conductivities $K_\perp$ and $K_\parallel$ - the reader can refer to Ref.[17] for details. The pump beam is expressed as[16,17]

$$P(k) = A_p \exp\left(\frac{-\pi^2 k^2 \omega_p^2}{2}\right) \tag{2}$$

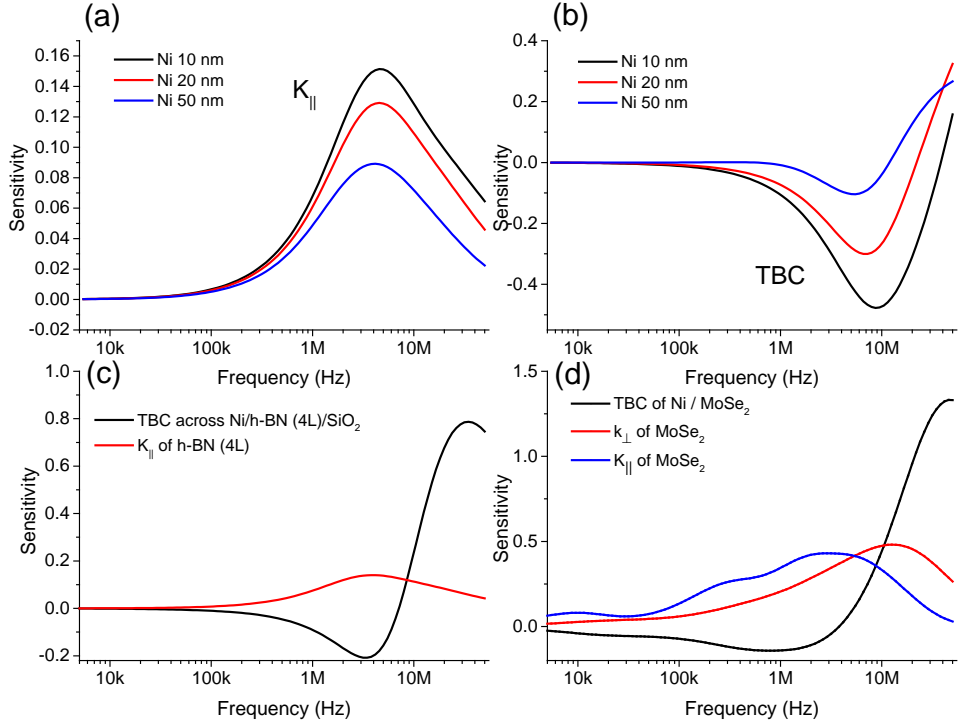

Figure 2 - (a) and (b) plot the sensitivity of the measured thermal phase to the in-plane thermal conductivity of single-layer graphene and the TBC across the Ni/graphene/SiO$_2$ interfaces for several values of the transducer thickness. (c) and (d) show the sensitivity to the measured thermal phase to different parameters using a 20 nm Ni transducer for 4-layer h-BN and MoSe$_2$ crystal, respectively.

where $A_p$ and $\omega_p$ are the total intensity and spot size of the pump beam. In the case of coaxial pump and probe beams the expression for $S(k)$ is equivalent to $P(k)$. For the case where the probe beam is offset with respect to the pump, S(k) is defined as[17]

$$S(k) = \frac{A_s}{\pi} \exp\left\{-\left[\left(\frac{\sqrt{2}x_0}{\omega_s}\right)^2 + \left(\frac{\pi\omega_s k}{\sqrt{2}}\right)^2\right]\right\} \sum_{n=0}^{\infty} \frac{1}{(n!)^2} \left(\frac{\sqrt{2}x_0}{\omega_s}\right)^{2n} l_n\left(\frac{\omega_s k}{\sqrt{2}}\right) \quad (3)$$

where, $x_0$ is the offset between pump and probe beams, $\omega_s$ is the probe spot size and $l_n(x)$ is defined recursively as

$$l_{n+1}(x) = -\frac{1}{x}\left[(\pi^2 x^3 - x)l_n(x) + \left(\frac{1}{4\pi^2} - x^2\right)l'_n(x) + \frac{x}{4\pi^2}l''_n(x)\right] \quad (4)$$

where $l'_n(x)$ and $l''_n(x)$ are the first and second derivative, respectively of $l_n(x)$, and $l_0 = \pi$.

To determine how sensitive the measured thermal phase $\theta$ (the phase of the complex temperature in eq. 1) is to different thermophysical parameters, the sensitivity to parameter $x$ is defined as

$$S_x = \frac{d\theta}{d\ln x} \tag{5}$$

We first consider the ability to sense single-layer graphene on a SiO$_2$/Si substrate. Figure 2(a) and (b) illustrate how a thin transducer layer enhances the sensitivity to measuring the in-plane thermal conductivity of single-layer graphene and the effective TBC across the metallic layer/graphene/SiO$_2$ interfaces, respectively. The distinct spectral sensitivity of various parameters, as also shown for example in figures 2(c) and (d) for the in-plane thermal conductivity of 4-layer h-BN and the TBC across the Ni/h-BN/SiO$_2$ interfaces, or the anisotropic thermal conductivity of MoSe$_2$ and TBC of the Ni/MoSe$_2$ interface allow us to determine these parameters concurrently. The sensitivity values obtained here compare very favorably with respect to other typical thermoreflectance approaches, where for example the sensitivity to $K_\parallel$ for graphene on SiO$_2$ is below 0.01 even when using smaller spot sizes[25] and 0.1 using beam offset approaches[21].

**Sample preparation and materials characterization**

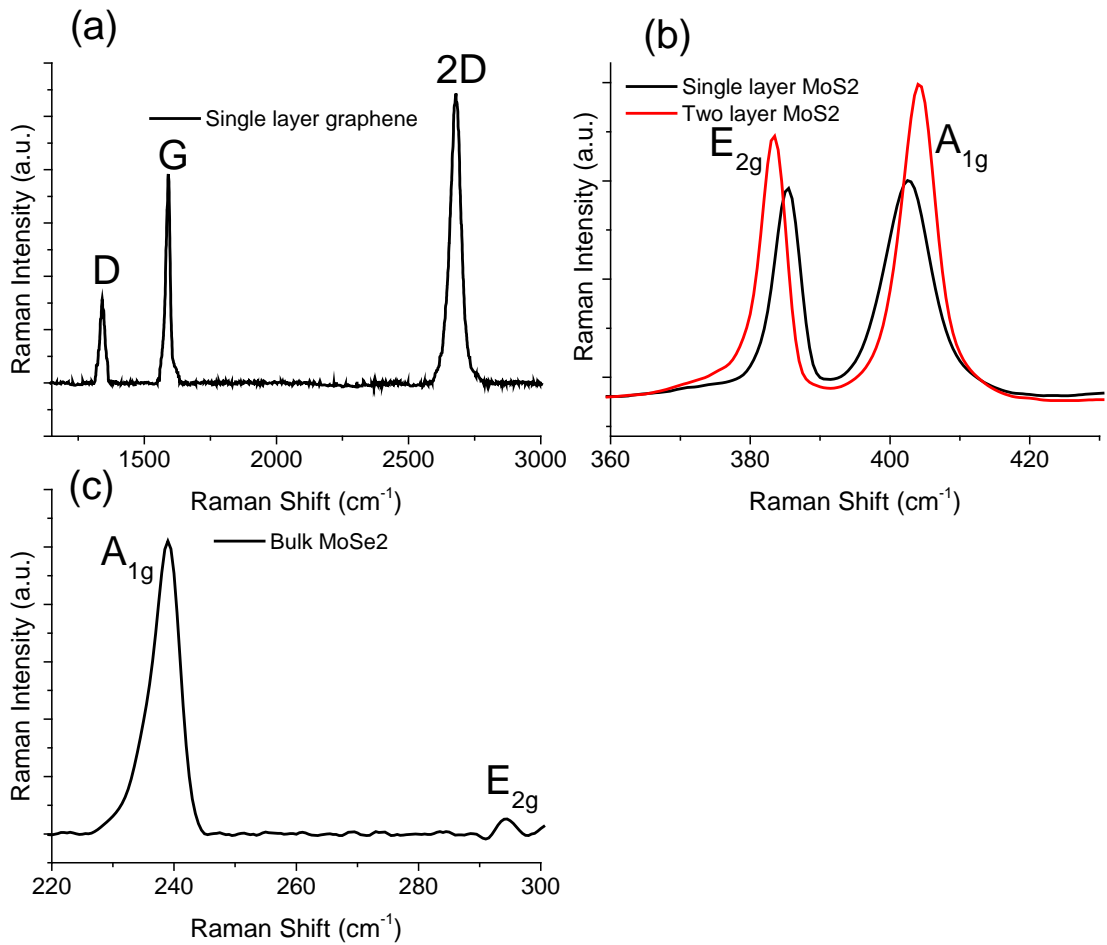

Figure 3 - Measured Raman Spectra of the 2D materials in this study.

Graphene, h-BN, MoS$_2$, and MoSe$_2$ samples were obtained from 2D Material, whereas Si and Al$_2$O$_3$ from MTI. Graphene, h-BN and MoS$_2$ were few-layer polycrystalline films synthesized through chemical vapor deposition and subsequent transfer to SiO$_2$(300 nm)/Si substrates. All samples were coated with a 20 nm Nickel film capped by 3 nm of Aluminum by sputter deposition. The Al layer prevents oxidation of the Ni layer. White light interferometry was used to measure the layer thickness. Electrical conductivity measurements were done by a 4-point probe to calculate the thermal conductivity of the metal layer via the Wiedemann-Franz law. A clean surface of the MoSe$_2$ crystal was obtained by exfoliation with adhesive tape.

We measured the Raman spectra for the single-layer graphene, few-layer h-BN, single and two-layer MoS$_2$ samples on SiO$_2$/Si substrate and bulk MoSe$_2$ using a Bruker Senterra dispersive Raman microscope at 532 nm. A 50X objective is used with a range of power from 5 mW to 10 mW. Figure 3(a) illustrates the Raman spectra of single-layer graphene where the G peak and 2D peaks are observed near 1591 cm$^{-1}$ and 2679 cm$^{-1}$ respectively. The observed D peak indicates the presence of defects that can be associated with grain boundaries in a polycrystalline sample. The 2D peak is a sharp Lorentzian peak with a full width at half maximum of 44 cm$^{-1}$, and it is more intense than the G peak, identifying single-layer graphene[29,30]. The comparison of the single and two-layer MoS$_2$ is shown in figure 3(b), where the in-plane and out-of-plane mode peaks for the two-layer sample are at 383.5 cm$^{-1}$ and 404.5 cm$^{-1}$, respectively, which upshift and downshift, respectively, for the single-layer[31–33]. In figure 3(c), the out-of-plane vibrational mode for bulk $MoSe_2$ is observed at 240 cm$^{-1}$, whereas the weaker in-plane vibration mode is found at 295 cm$^{-1}$ [31]. In the case of h-BN excited with a laser in the visible range, the Raman processes are non-resonant and consequently, Raman spectra are much weaker[34]. The number of layers for the two of the h-BN samples was estimated through AFM characterization by measuring the step height with respect to the $SiO_2$/Si substrate and found to be 4 layers (~2nm) and 20 layers (~9nm).

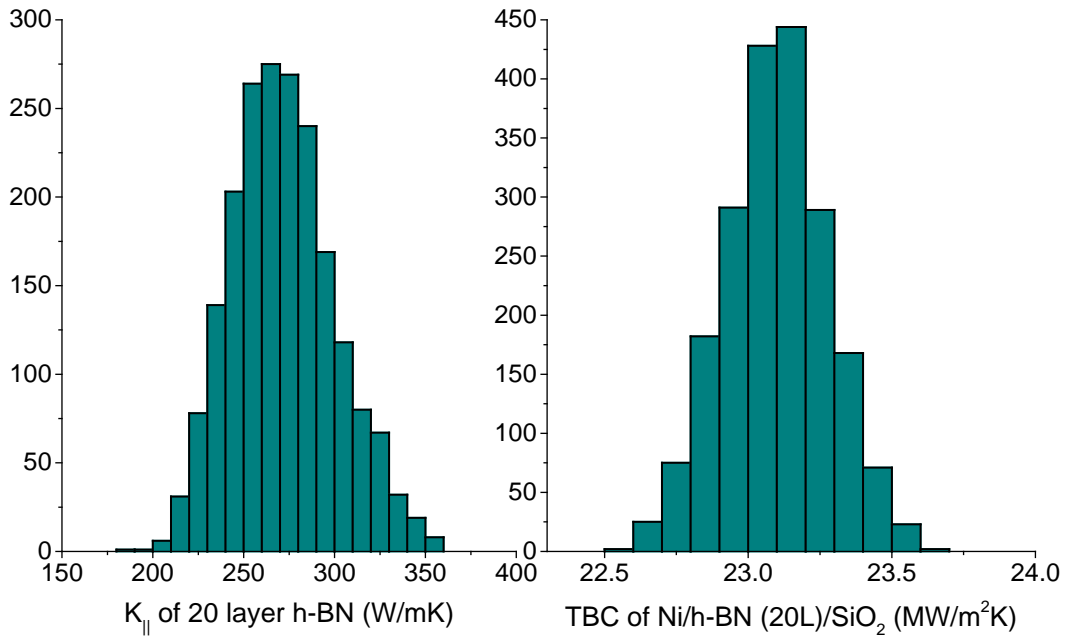

Figure 4 - Histograms for the derived thermal properties of the 20-layer h-BN sample, obtained by a Monte Carlo model of the uncertainties in the assumed parameters.

**Uncertainty estimation**

In spite of the optical spot size measurements we perform, small errors affect the derived thermal properties of single and few-layer materials more than for the case of single crystal samples. Additionally, errors in the measurements of the Ni film thickness and its electrical conductivity, or deviations in any of the other parameters from their true value can increase the uncertainty. We estimate the propagation of these errors by a Monte Carlo approach, where random errors are introduced in the assumed parameters to estimate the variation on the derived properties. The standard deviation in optical spot size is estimated from several independent measurements to be 3%. Additionally, the uncertainty in the measured phase originating from experimental noise of 0.1 degree is also included. 2,000 Monte Carlo runs were used to verify that the results converged. For each Monte Carlo run both the fitted parameters and their uncertainties from the goodness of fit were recorded, the latter was used to weigh the value of its associated parameter for statistical significance. The aggregate results were obtained by calculating the mean value and the standard deviation of each derived parameter due to uncertainties in the values assumed. As an example, the resulting uncertainty histograms of the measured in-plane thermal conductivity and effective conductance of Ni/graphene/SiO$_2$ for 20-layer h-BN sample is shown in Figure 4. The confidence of the results could be increased by performing several independent measurements[19].

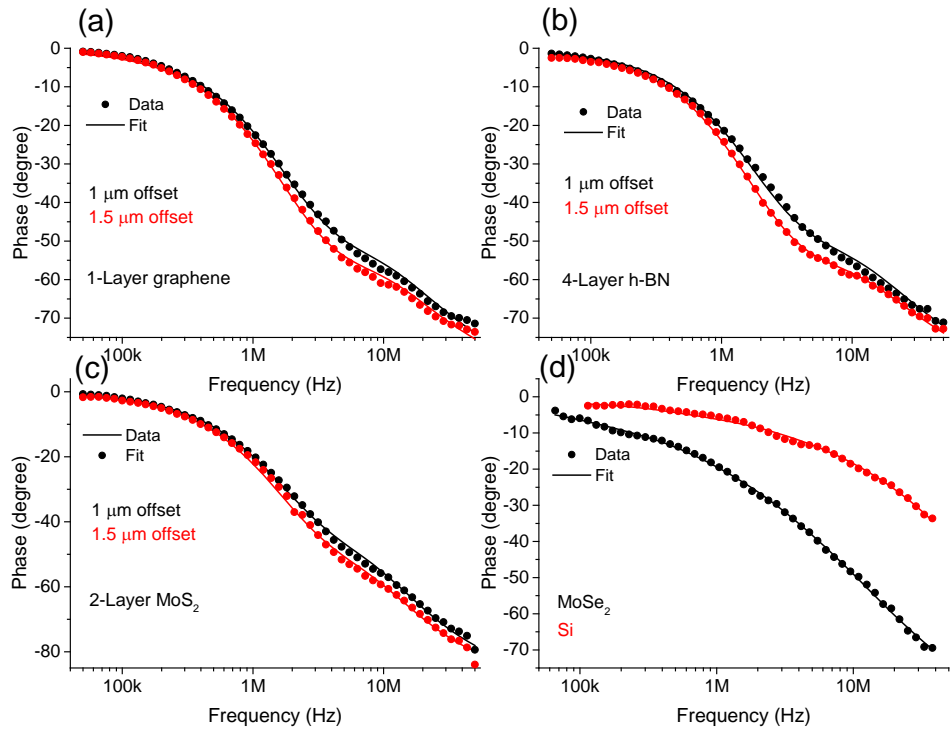

Figure 5 - Panels (a), (b), and (c) show aggregate data (symbols) and global fit (solid lines) using 2 beam offset values: 1 µm (black) and 1.5 µm (red) for the Ni/graphene(1L)/$SiO_2$/Si, Ni/h-BN (4L)/$SiO_2$/Si, and Ni/$MoS_2$ (2L)/$SiO_2$/Si samples, respectively. (d) data and fit at 0 µm offset for the Ni/$MoSe_2$ crystal (black) and Ni/native $SiO_2$/Si (red) samples, respectively.

**Results and discussions**

We have measured the anisotropic thermal conductivity as well as effective thermal boundary conductance (TBC) between the Ni layer and several sample/substrate combinations by modulating over a wide frequency range from 50 KHz to 50 MHz through concentric and beam offset frequency domain magneto-optical Kerr effect. Representative FD-MOKE data with fits are presented in figure 5, and all the parameters fitted from these measurements are presented in Table I and Figure 6. Figure 6 compares the measurements obtained in this work with literature data.

Table 1 – Thermal conductivity and TBC for all samples measured in this work. The TBC is for the Ni/crystal interface for bulk crystals or across the Ni/sample/SiO$_2$ structure for layered crystals.

| Sample | Type | $K_\perp$ (W/mK) | $K_\parallel$ (W/mK) | TBC (MW/m$^2$K) |
| --- | --- | --- | --- | --- |
| Silicon | Bulk | 123 ± 2 | 123 ± 2 | 134 ± 3 |
| c-Sapphire | Bulk | 34 ± 2 | 34 ± 2 | 180 ± 20 |
| MoSe$_2$ | Bulk | 1 ± 0.1 | 30 ± 2 | 24 ± 1 |
| Graphene | 1-Layer | - | 636 ± 140 | 17 ± 0.2 |
| h-BN | 4-Layer | - | 242 ± 22 | 19.6 ± 0.3 |
| h-BN | 20-Layer | - | 270 ± 28 | 23.1 ± 0.2 |
| MoS$_2$ | 1-Layer | - | 63 ± 22 | 15 ± 0.2 |
| MoS$_2$ | 2-Layer | - | 74 ± 10 | 13 ± 0.2 |

Reference measurements were carried out on two standard crystal samples, Silicon and Sapphire, to verify the validity of the FD-MOKE technique. For Silicon (which includes its native oxide layer) we obtained a thermal conductivity of 123 ± 2 W/mK and a TBC with Ni of 134 ± 3 MW/m$^2$K. The value for the TBC of Ni/Si is expected to be similar to that of the Al/Si interface, as one would expect from the similar phonon dispersion properties of Al and Ni[35], and it is indeed similar to the TBC of the Al/Si interface of 116 MW/m$^2$K reported elswhere[20]. The measured thermal conductivity of Si is ~15% lower than that expected for the bulk intrinsic crystal, and this is because these measurements were performed with a small pump/probe optical spot size, which is known to manifest itself in a reduced apparent thermal conductivity due to the onset of non-diffusive heat transport[36] in crystals having very long mean free path heat carriers. The result obtained here is in line with the non-diffusive heat transport reported by Wilson and Cahill at comparable spot sizes[36]. The presence of the native oxide is expected to cause the reduction in the apparent thermal conductivity to be more sensitive to spot size rather than modulation frequency, given our experimental conditions[37]. For Sapphire the thermal conductivity measured is 34 ± 2 W/mK, is in agreement with literature values and non-diffusive transport is not detected given the shorter phonon mean free paths in Sapphire. The weak anisotropy (<10%) expected along the *c*-axis is not captured well within the sensitivity of this measurement. The TBC at the Ni/Sapphire interface of 180 ± 20 MW/m²K is lower than expected, given that the Al/Sapphire TBC has been measured to be near 250 MW/m²K [35,38]. Although our value is comparable to that of other reports[19,20], the reduced value observed here may be affected by adsorbates on the Sapphire surface prior to the Ni deposition.

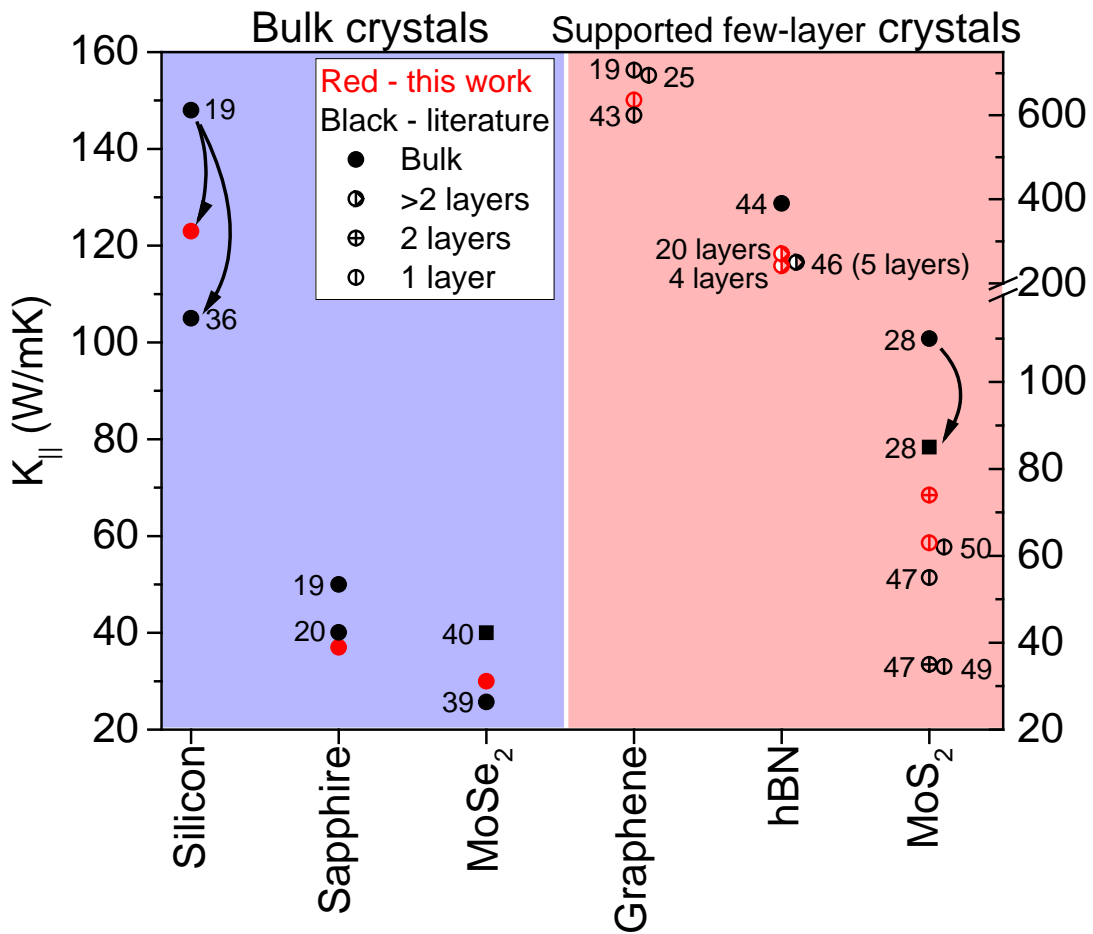

Figure 6 - Comparison of in-plane thermal conductivity values at room temperature: black symbols are for values reported in the literature along with corresponding reference number, red symbols are measurements reported in this work, circles refer to experimental values and squares to theory. The symbol's fill distinguishes between bulk values and few-layer crystals. Literature values for few-layer crystals are supported samples. The arrows symbolize the reduction from bulk values due to non-diffusive heat transport, which leads to a lower apparent thermal conductivity.

We now report results obtained for a bulk 2D crystal. As shown in figure 2(d), we have enough sensitivity to the anisotropic thermal conductivity in MoSe$_2$ as well as TBC of the Ni/MoSe$_2$ interface, and hence we can measure these three parameters concurrently. The experimental data with the analytical fitted solution is in figure 5(d) and the fit yields the out-of-plane and in-plane thermal conductivity of MoSe$_2$: $K_\perp$ = 1 ± 0.1 W/mK and $K_\parallel$ = 30 ± 2 W/mK respectively, and the TBC of Ni/MoSe$_2$ of 24 ± 1 MW/m$^2$K. The measured thermal conductivity is highly anisotropic as expected from the weaker atomic bonding along the *c* axis, and is in agreement with the reported in-plane thermal conductivity of a 80 nm thick MoSe$_2$ film (25.7 ± 7.7 W/mK)[39]. We are not aware of experimental reports made on bulk crystals. Our data can be compared to first-principles calculations of bulk MoSe$_2$[40], reported to be $K_\perp$ = 3.5 W/mK and $K_\parallel$ = 40 W/mK respectively. The discrepancy with respect to theory likely originates from extrinsic effects not included in the first-principles calculations[40] such as crystal defects and boundary scattering. The measured thermal boundary conductance of 24 ± 1 MW/m$^2$K is very low compared to typical metal dielectric interfaces, but this is common for layered crystals, and in good agreement with the measurement for MoS$_2$[28]. The low thermal boundary conductance has been ascribed to the effect of phonon focusing in crystals with elastic anisotropy[41].

We now move to measurements made on few-layer 2D materials. For these cases, the determination of the in-plane thermal properties is challenging, particularly for materials with relatively small in-plane thermal conductivity, due to the low sensitivity to in-plane heat transport. Measurements on graphene have been reported before[19,25], and the approach with FD-MOKE presented here can serve to verify the validity for measurements on these ultra-thin layers. Here, the use of a thin transducer enhances the sensitivity to lateral heat transport. As seen in figure 2(a) and (b) the experiment has enhanced sensitivity to the in-plane thermal conductivity of single-layer graphene and effective thermal boundary conductance across the Ni/graphene/SiO$_2$ structure. We model the graphene layer as having a thickness of 0.335 nm and negligible out-of-plane thermal resistance and treat this together with the TBC of Ni/graphene and the TBC of graphene/SiO$_2$. We performed measurements at several pump-probe beam offsets, and the fitted value for $K_\parallel$ of graphene is 636 ± 140 W/mK and the TBC across Ni/graphene/SiO$_2$ is 17 ± 0.2 MW/m$^2$K. The measured in-plane thermal conductivity is in good agreement with our previous measurement of supported graphene through beam offset FDTR[19] and with other literature values[25,42,43], though previous measurements were done using Al or Ti transducers. This further strengthens Yang's observation that the in-plane thermal conductivity of graphene is independent of the metal contact[25]. The reduction of the in plane thermal conductivity in a SiO$_2$-supported geometry as opposed to a suspended geometry is due to boundary scattering from the SiO$_2$, and it is expected that the thermal conductivity could be enhanced by replacing the SiO$_2$ support with h-BN.

We now move to few-layer h-BN. As shown in figure 2(c), the measurement sensitivity to both in-plane thermal conductivity and effective TBC across the Ni/h-BN/SiO$_2$ interfaces is comparable to the case of graphene. We perform these measurements for a 4-layer h-BN film by performing measurements with several beam offset values (figure 5(b)). We model the 4-layer h-BN as an interface having a thickness of 1.33 nm and negligible out-of-plane thermal resistance and treat this together with the TBC of Ni/h-BN and the TBC of h-BN/SiO$_2$. The fit yields $K_{\parallel}$ of h-BN to be 242 ± 22 W/mK, and the TBC across Ni/h-BN/SiO$_2$ to be 19.6 ± 0.3 MW/m$^2$K. We also measured the thermal properties of a 20-layer h-BN film with several beam offsets and obtained $K_{\parallel}$ of 270 ± 28 W/mK, and TBC across Ni/h-BN/SiO$_2$ of 23.1 ± 0.2 MW/m$^2$K. The measured thermal conductivity is lower than the reported bulk value of 390 W/mK[44] at room temperature but in line with the values reported of 250 W/mK for 5-layers[45], 227–280 W/mK for 9 layers[46], and 360 W/mK for 11-layers[45]. Similarly to these reports, we also find an increasing trend for the in-plane thermal conductivity of h-BN with the number of layers. The dependence with sample thickness has been ascribed to scattering with phonon modes in the substrate similarly to graphene, which is reduced for thicker films.

For the case of MoS$_2$, the thermal conductivity of the bulk crystal was measured to be between 85–110 W/mK as a function of laser spot size through TR-MOKE[28]. However, there is a large discrepancy in the reported values for single and two-layer MoS$_2$, particularly wen comparing suspended or supported layers, and there is no apparent systematic trend for the reported thermal conductivity as function of the number of layers. We performed measurements on single and two-layer MoS$_2$ (figure 5(c)) for several beam offset values. The fit gives $K_{\parallel}$ of 63 ± 22 W/mK and TBC across Ni/MoS$_2$/SiO$_2$ of 15 ± 0.2 MW/m$^2$K for single-layer MoS$_2$ and $K_{\parallel}$ of 74 ± 10 W/mK and TBC of 13 ± 0.2 MW/m$^2$K for the two-layer MoS$_2$. The relatively large uncertainty in the single-layer film is due to the reduced in-plane heat flux in this structure, lowering the sensitivity of the measurement. The trend of $K_{\parallel}$ with thickness is opposite to the values of 55 W/mK for single-layer MoS$_2$ and 35 W/mK for two-layer MoS$_2$ measured by Raman optothermal technique[47] and the prediction from first-principles calculations[48]. The apparent discrepancies may be due to larger error in the value obtained for single-layer MoS$_2$, or scattering induced by the supporting SiO$_2$, as the predicted values as function of thickness is obtained for suspended samples.

**Conclusions**

We have implemented a new technique, frequency domain magneto-optical Kerr effect to enhance the sensitivity to lateral heat transport and measure the thermal conductivity of anisotropic materials including atom-thick materials. Using this technique, we measured the anisotropic thermal conductivity of a wide range of 2D materials which are interest for novel device applications. We performed the first experimental study on bulk MoSe$_2$ crystal, and our measured value is in good agreement with recently reported nm thick MoSe$_2$ film. Due to the enhancement in experimental sensitivity, we are also able to perform measurements on single, and few-layer 2D materials. Overall, this technique will allow more robust measurements on 2D layered materials including one-atom thick films on various supports, providing a better understanding of the heat dissipation problems in the thermal management of nanoelectronic and optoelectronic devices based on these materials.

**Acknowledgments**

We would like to acknowledge the Natural Sciences and Engineering Research Council of Canada, CMC Microsystems and York University for financial support.**References**

1. Novoselov, K. S., Mishchenko, A., Carvalho, A. & Neto, A. H. C. 2D materials and van der Waals heterostructures. *Science* **353**, aac9439 (2016).
2. Geim, A. K. & Grigorieva, I. V. Van der Waals heterostructures. *Nature* **499**, 419 (2013).
3. Novoselov, K. S. *et al.* A roadmap for graphene. *Nature* **490**, 192 (2012).
4. Withers, F. *et al.* Light-emitting diodes by band-structure engineering in van der Waals heterostructures. *Nature Materials* **14**, 301 (2015).
5. Withers, F. *et al.* Heterostructures Produced from Nanosheet-Based Inks. *Nano Letters* **14**, 3987 (2014).
6. Yankowitz, M., Xue, J. & LeRoy, B. J. Graphene on hexagonal boron nitride. *J. Phys.: Condens. Matter* **26**, 303201 (2014).
7. Zhang, Z., Hu, S., Chen, J. & Li, B. Hexagonal boron nitride: a promising substrate for graphene with high heat dissipation. *Nanotechnology* **28**, 225704 (2017).
8. Choi, W. *et al.* Recent development of two-dimensional transition metal dichalcogenides and their applications. *Materials Today* **20**, 116 (2017).
9. Tian, T., Rice, P., Santos, E. J. G. & Shih, C.-J. Multiscale Analysis for Field-Effect Penetration through Two-Dimensional Materials. *Nano Lett.* **16**, 5044 (2016).
10. Mak, K. F., Lee, C., Hone, J., Shan, J. & Heinz, T. F. Atomically Thin ${\mathrm{MoS}}_{2}$: A New Direct-Gap Semiconductor. *Phys. Rev. Lett.* **105**, 136805 (2010).
11. Zhao, Y., Wang, W., Li, C. & He, L. First-principles study of nonmetal doped monolayer MoSe 2 for tunable electronic and photocatalytic properties. *Scientific Reports* **7**, 17088 (2017).
12. Jung, C. *et al.* Highly Crystalline CVD-grown Multilayer MoSe$_2$ Thin Film Transistor for Fast Photodetector. *Scientific Reports* **5**, 15313 (2015).
13. Shim, G. W. *et al.* Large-Area Single-Layer MoSe2 and Its van der Waals Heterostructures. *ACS Nano* **8**, 6655 (2014).
14. Peimyoo, N. *et al.* Thermal conductivity determination of suspended mono- and bilayer WS2 by Raman spectroscopy. *Nano Res.* **8**, 1210 (2015).
15. Wang, Y., Xu, N., Li, D. & Zhu, J. Thermal Properties of Two Dimensional Layered Materials. *Advanced Functional Materials* **27**, 1604134 (2017).
16. Cahill, D. G. Analysis of heat flow in layered structures for time-domain thermoreflectance. *Review of Scientific Instruments* **75**, 5119 (2004).
17. Feser, J. P. & Cahill, D. G. Probing anisotropic heat transport using time-domain thermoreflectance with offset laser spots. *Review of Scientific Instruments* **83**, 104901 (2012).
18. Schmidt, A. J., Chen, X. & Chen, G. Pulse accumulation, radial heat conduction, and anisotropic thermal conductivity in pump-probe transient thermoreflectance. *Review of Scientific Instruments* **79**, 114902 (2008).
19. Rahman, M. *et al.* Measuring the thermal properties of anisotropic materials using beam-offset frequency domain thermoreflectance. *Journal of Applied Physics* **123**, 245110 (2018).
14